\documentclass[a4paper,10pt,twocolumn,aps,showpacs,prb]{revtex4-1}
\usepackage{graphicx}% Include figure files
\usepackage{dcolumn}% Align table columns on decimal point
\usepackage{bm}% bold math
\usepackage{slashed}
\usepackage{amssymb}
\usepackage{amsmath}
\usepackage{amsthm}
\usepackage{color}
\usepackage{esint}

\begin{document}

\title{Interaction Protected Topological Insulators with Time Reversal Symmetry}

\author{Raul A. Santos$^{1,2}$  and D.B.  Gutman$^{1}$}
%\email{raul.santosza@biu.ac.il} 
\affiliation{$^{1}$Department of Physics, Bar-Ilan University, Ramat Gan, 52900, Israel}
\affiliation{$^{2}$Department of Condensed Matter Physics, Weizmann Institute of Science, Rehovot 76100, Israel}

%\author{Dmitri Gutman}
%\email{Dmitri.Gutman@biu.ac.il}
%\affiliation{Department of Physics, Bar-Ilan University, Ramat Gan, 52900, Israel}

\begin{abstract}
Anderson's localization on the edge of two dimensional time reversal (TR) topological insulator (TI) is studied.  
For the non-interacting  case the topological protection  acts  accordingly to the $\mathbb{Z}_2$ classification,  
leading to conducting and insulating phases for odd  and  even fillings respectively.
In the  presence of repulsive interaction  the phase diagram is notably changed. 
We show that for sufficiently strong values of the interaction the zero temperature fixed point  
of the TI  is conducting,  including the  case of even fillings. We compute the boundaries 
of the conducting phase for various fillings and types of disorder. 
\end{abstract}

\preprint{}

\maketitle

\section{Introduction}

Time reversal  non-interacting TIs are realized in materials with strong spin orbit interaction in two \cite{Konig2007,Konig2008,Drozdov} and three
\cite{Hsieh2008,Hsieh2009,Xia2009,Hsieh2009b,Zhang2009} dimensions.
Also known as quantum spin Hall insulators, 
these materials have an insulating bulk, while hosting gapless surface states.

For the non-interacting  case the classification of disordered  TIs is complete \cite{Schnyder2008,Kitaev2009,Ludwig2010,Shinsei2010,Qi2011}.
In two dimensions, non interacting TR invariant TIs are classified by a $\mathbb{Z}_2$ topological
invariant, according to  the number of helical edge states \cite{Kane2005,Bernevig2006}. 
The back scattering  by TR invariant disorder is possible only between states  that are not  Kramer's
partners. Therefore for  an odd number of the helical states the conducting state is protected.
For an  even number of helical edge modes the electrons can be localized completely 
by scattering among  non-Kramers pairs. This  state is therefore  equivalent to a trivial insulator
in agreement  with a more general Haldane criterion \cite{Haldane1995}.

Interacting  integer and fractional TI's  are  a subject of active research.
Interaction may  lead to strongly correlated ground states \cite{Levin2009,Neupert2014,Klinovaja2014b,LSantos2011,Santos2015}  
with fractional excitations and  non trivial statistics \cite{Neupert2011,Scharfenberger2011,Levin2012}.  
The classification of TI's  in the interacting case is yet unknown. 
An approach,  based on thermal response and its relation to a quantum anomaly,  
valid beyond single particle picture, was proposed in Ref. \cite{Ludwig2012}. 

It is commonly accepted, that the charge transport in the ideal  TI occurs  via protected a single helical edge mode, with 
the universal quantized conductance $2e^2/h$. In reality  the conductance differ from this  value 
due to  back scattering processes.  The latter  may occur via the combination of the 
two electron scattering and the disorder potential \cite{Stephan,Kainaris2014}, coupling to the bulk via electron 
puddles\cite{Glazman_Gefen2014} or due to the magnetic impurities. 
In the later case  the interaction stabilizes the conducting phase,
and the quantum phase transition as function of Luttinger liquid (LL) parameter is predicted 
for the fractional TIs \cite{Beri2012}. 

In this work we study the localization by TR disorder on the edge states of a TI 
in the presence of repulsive interaction.
Although we  focus on the TI's at integer fillings ($\nu$),  that  in the absence of the disorder possess
$\nu$ helical edge states, the same analysis applies for  narrow stripe of  TI at $\nu =1$.
\cite{Chia-Wei}
The inclusion of TR disorder drives the non-interacting system to a state with 
a single or no helical edge states.
We show that  the presence  of the repulsive interaction can stabilize the conducting phase.

We model the disorder by a  short range static  potential that 
due to the spin orbit interaction mixes  different helical states, except those that are  connected by the TR symmetry. 
We consider a generic finite range interaction between the electrons, with all 
possible matrix elements allowed by symmetry. 
We consider the case of a single impurity and the random disorder, with a  scattering length shorter that the sample size.
We  perform one loop renormalization group (RG) analysis,
analogous to Kane-Fisher \cite{Kane1992} and  Giamarchi-Schultz \cite{Giamarchi1988} study  of localization in 
one dimensional systems.
 
Our analysis shows that the low energy fixed point is determined 
by the magnitude of the interaction and  its effective radius. 
For interaction stronger that some critical value  the  low temperature phase is conducting.   

\section{The Model}

The appearance  of the helical edge states can be understood  on the level of  non interacting electrons.
In the presence of a Rashba spin-orbit (SO) interaction the single particle Hamiltonian is given by
\begin{eqnarray}\label{Ham_Rashba}
 H&=&\frac{\hat{p}_x^2+\hat{p_y}^2}{2m_e}+\alpha_{SO}(\vec{p}\times\vec{\sigma})\cdot\nabla V(x,y)+ V(x,y),
\end{eqnarray}
%TR TIs are expected to appear in systems with large spin-orbit (SO) coupling. The SO coupling induces an effective spin
%dependent \textquotedblleft magnetic\textquotedblright field with opposite orientation for different spin projections  \cite{Kane2005,Bernevig2006}.
%A natural single particle Hamiltonian is
where  $m_e$ is effective mass of an electron and $\alpha_{SO}$ is the strength of the SO coupling.
For the parabolic potential $V(y)={y^2}/{2m_e\alpha_{SO}^2}$, the Hamiltonian (\ref{Ham_Rashba})
\begin{equation}
 H=\frac{\hat{p_y}^2}{2m_e}+\frac{1}{2m_e}\left(\hat{p}_x-\frac{y}{\alpha_{SO}}\sigma_z\right)^2
\end{equation}
 corresponds to two replicas of fermions subject to opposite magnetic fields  
$|B|=\alpha_{SO}^{-1}$.
For the integer fillings,  $\nu=\alpha_{SO}^{-1}A/\Phi_0$ ($A$ being an area of sample and $\Phi_0=hc/e$ the flux quantum)
 the bulk forms an incompressible state with  a  gap of size $\hbar\alpha_{SO}^{-1}/m_e$.

\begin{figure}
\includegraphics[width=1\columnwidth]{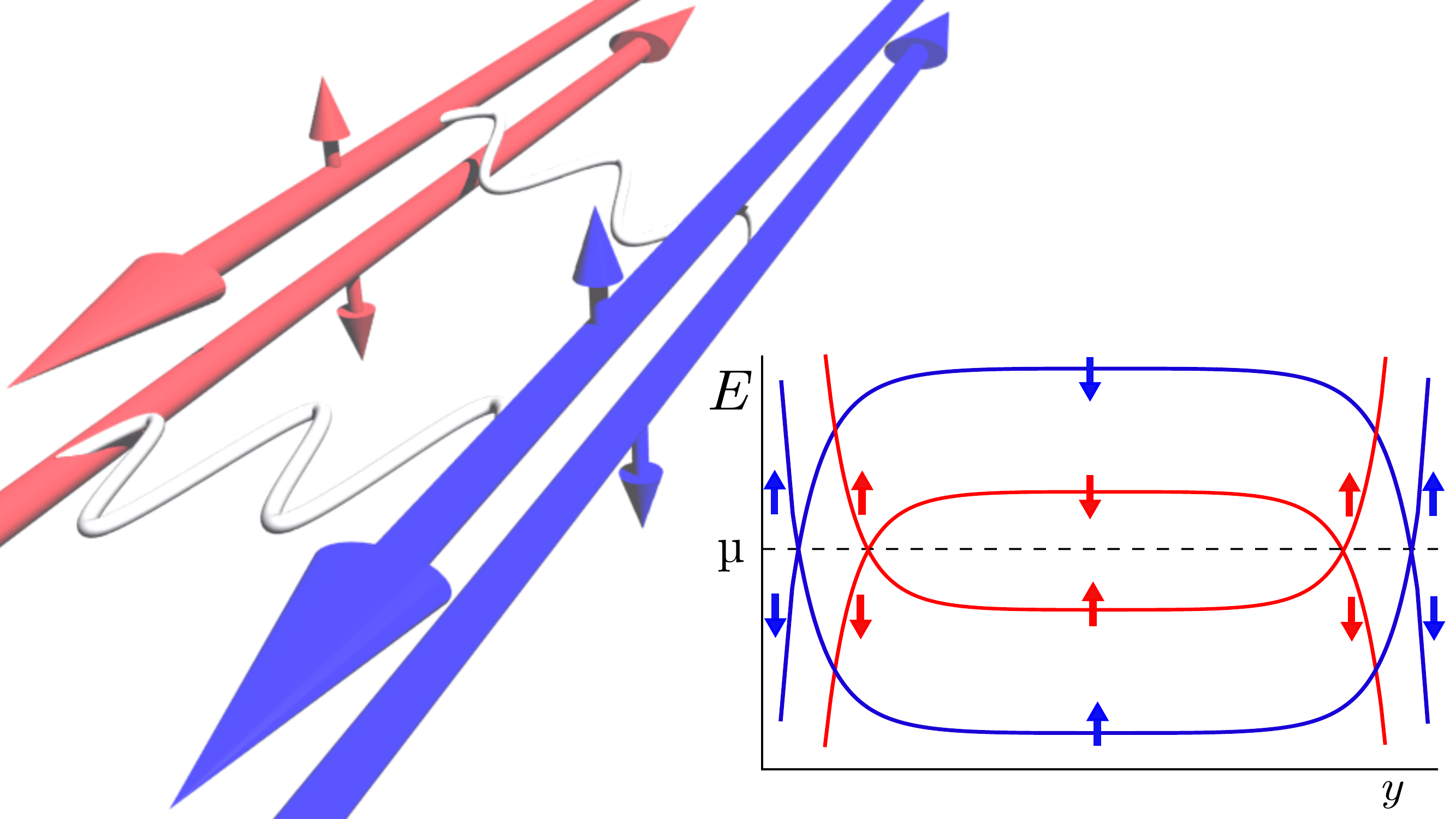}
\caption{\label{fig:wires}(color on line)  Helical edge modes. The wavy lines represent  the
back-scattering events  between non-Kramers pairs  (in red and blue). Insert.- Schematic band structure of a TI.}
\end{figure}

An addition of a smooth confining potential curves the Landau levels, as shown in Fig.\ref{fig:wires},
leading to $\nu$  gapless helical edge states \cite{Kane2005,Bernevig2006}.
Assuming  that the single particle gap formed in the bulk is not closed  the effects of interaction 
can be taken into account  within the helical edge states.
This phenomenological approach  can be microscopically justified within  
the sliding Luttinger liquid model \cite{Klinovaja2014b,Santos2015}.
However,  the resulting helical edge description  is believed to be the correct low energy model,
valid beyond the sliding LL approximation.

To account for the interaction,  it is natural to pass to the bosonic description,
defining   bosonic fields $R_{i}/L_{i}$,
related to the  right/left density components  by $\rho_{R,i}=\partial_x R_i/2\pi$ and $\rho_{L,i}=-\partial_x L_i/2\pi$ \cite{GiamarchiBook2003}.
 These fields satisfy  the canonic commutation relations $[R_{j}(x),R_{j'}(y)]=-[L_{j}(x),L_{j'}(y)]=i\pi{\rm sgn}(x-y)\delta_{jj'}$.
The electronic operators are represented as
$\psi_{R,j}= e^{-iR_j}/\sqrt{2\pi a},\quad\mbox{and}\quad\psi_{L,j} = e^{-iL_i}/\sqrt{2\pi a},$ with $a$ a short
distance cut-off.

In  the absence of the  Umklapp  and $2k_F$ electron-electron scattering  the interaction between $\nu$ modes, consistent with TR symmetry,  is represented by the following action
\begin{equation}\label{action}
 S=\frac{1}{2\pi}\int dxdt\left(\partial_x \boldsymbol{\Phi}^T\mathcal{K}\partial_t\boldsymbol{\Phi}-\partial_x\boldsymbol{\Phi}^T\mathcal{M}\partial_x\boldsymbol{\Phi}\right),
\end{equation}
where we use the compact notations  
$\boldsymbol{\Phi}=(\boldsymbol{R},\boldsymbol{L})$, 
$\boldsymbol{R}=(R_1,R_2, \dots, R_\nu)$ and similarly for $\boldsymbol{L}$. Here the matrix $\mathcal{K}$ encodes
the commutation relations of the fields and can be written as $\mathcal{K}=\sigma_z\otimes\mathbb{I}_\nu$, where
the $\sigma_z$ is a Pauli matrix that acts in the right/left movers 
subspace while $\mathbb{I}_\nu$ is the identity matrix in the space of $\nu$ modes.
The  positive definite matrix $\mathcal{M}$  accounts for interaction. The helical edge modes are separated in space by a distance $d$.
 
The symmetry under TR requires that  $\{T,\mathcal{K}\}=[T,\mathcal{M}]=0$, where
$T=\sigma_x\otimes\mathbb{I}_\nu$ is a time reversal symmetry operator. 
This restricts the interaction matrix  $\mathcal{M}$ to the form
\begin{equation}\label{matrix}
 \mathcal{M}=\begin{bmatrix}
 M_{\rm fw} & M_{\rm bw}\\
 M_{\rm bw} & M_{\rm fw} 
 \end{bmatrix}=\mathbb{I}_2\otimes M_{\rm fw}+\sigma_x\otimes M_{\rm bw},
\end{equation}

\noindent where  $(M_{\rm fw})_{ij}$ describes  the {\it forward} interaction  between the copropagating 
modes $\rho_{R,i}$ and $\rho_{R,j}$ (similarly for left movers). 
%Similarly $M^{\perp}$ represents interactions between opposite spins. 
$(M_{\rm bw})_{ij}$ is a an element of a symmetric matrix in the channel space that describes the {\it backward} interaction between 
$R_i$ and $L_j$.
We assume that the interaction between helical modes $i$ and $j$  is translationally invariant, 
and depends only on the relative  distance $|i-j|$.
In the absence of disorder the spectrum of this model is gapless.

The presence  of impurities may dramatically change the states of edge modes.
TR invariant disorder mixes helical states that belong  to  different Kramer's pairs
and induces  backward scattering processes. We consider  two cases: 
(a) single impurity scattering and
(b) random disorder. 
Single impurity scattering is the dominant process if the mean free path of the electrons is
larger than the sample size. 
In the opposite case (the electrons' mean free path is smaller than the sample size)
the localization  is  dominated by  multiple scattering.

\subsection{Single Impurity}

We analyze the single impurity case first.
The single impurity,  located on the edge  backscatter between 
states $i$ and $j$ that are not connected by TR symmetry,  $\mathcal{O}^{\rm imp}_{ij}=\mu_{ij}\psi_{R,i}^\dagger(0)\psi_{L,j}(0)$,
where $\mu_{ij}$ is proportional to the impurity potential at $k=2k_F$.
The renormalization of the strength $\mu_{ij}$  is a straightforward  generalization of standard Kane-Fisher analysis\cite{Kane1992} 
\begin{equation}\label{single_imp}
 \frac{d\mu_{ij}}{dl}=(1-\Delta_{ij})\mu_{ij}\,.
\end{equation}
Here $\Delta_{ij}$ is the scaling dimension of the scattering process 
$\langle\psi^\dagger_{R,i}(\tau)\psi_{L,j}(\tau)\psi^\dagger_{L,j}(0)\psi_{R,i}(0)\rangle\sim |\tau|^{-2\Delta_{ij}}$.
From Eq. (\ref{single_imp}) it follows that a single impurity is an irrelevant perturbation if 
 $\Delta_{ij}>1$.

The scaling dimension $\Delta$ is controlled by the backward interaction matrix
$M_{\rm bw}$. For the simple case $\nu=2$ only  two helical modes propagate on the edge.
If the separation  between the helical states is larger than the interaction radius
%$\Delta$ in the case of no interaction between different helical edge modes.
the effective Hamiltonian is given by $H=H_{\rm fw}+H_{\rm bw}$ with
\begin{eqnarray}\label{Ham_simple}\nonumber
 H_{\rm fw}&=&\frac{1}{2\pi}\sum_{i=1}^{2}\int dx \left(v_F+\frac{g_4^0}{\pi}\right)((\partial_x R_{i})^2+(\partial_x L_{i})^2),\\
 H_{\rm bw}&=&-\frac{g_2^0}{\pi^2}\sum_{i=1}^{2}\int dx\partial_x R_{i}\partial_x L_i.
\end{eqnarray}
Here $v_F$ the Fermi velocity while $g_4^0$ and $g_2^0$ parameterize the forward and backward interactions respectively.
Defining the fields $\varphi_i=(R_i-L_i)/\sqrt{2}$ and $\theta_i=(R_i+L_i)/\sqrt{2}$, the above Hamiltonian can be written in a Luttinger liquid (LL) form

\begin{equation}
 H=\frac{u}{2\pi}\sum_{i=1}^2\int dx \left((\partial_x \theta_i)^2 K +\frac{(\partial_x \varphi_i)^2 }{K}\right),
\end{equation}

\noindent where $u=(v_F+g_4^0/\pi)\sqrt{1-(\lambda_2)^2}$ is the renormalized sound velocity, $\lambda_2=\frac{g_2^0}{\pi(v_F+g_4^0/\pi)}$ and $K=\sqrt{\frac{1-\lambda_2}{1+\lambda_2}}$ is the LL parameter ($K<1$ for repulsive and $K>1$ for attractive interaction within the mode).  
In this case  the  single impurity operator $\mathcal{O}^{\rm imp}_{12}$ has the scaling dimension $\Delta_{12}=K/2+1/2K$, 
so the single impurity is irrelevant for any interaction. This is in stark contrast with the non helical  LL where 
the scaling dimension of the disorder operator is $K$, so the disorder is irrelevant only for an  attractive interaction \cite{Giamarchi1988}.
If  the  LL's  are different  the scaling
dimension is  controlled by both LL parameters $\Delta_{12}=(K_1+1/K_1+K_2+1/K_2)/4$. This result
implies  that the scattering between different edge states is accompanied 
by  the zero bias anomaly that suppresses the probability of this process. 
This is in  contrast to Kane-Fisher  case\cite{Kane1992}, 
where  the back scattering occurs in the same LL  and  has  no zero bias anomaly suppression. 
For TI  this  processes  is forbidden by the TR symmetry.

For the case where the helical edge states are located within the radius of interaction (or for any long range interaction potential)
the interaction matrices (\ref{matrix}) are 
\begin{equation}
M_{\rm fw}=\left(\begin{matrix}
v_F+g_4^0&g_4^1\\ g_4^1 &v_F+g_4^0
\end{matrix} \right), \,\,
M_{\rm bw}=-\left(\begin{matrix}
g_2^0& g_2^1\\ g_2^1&g_2^0 \end{matrix} \right).
\end{equation} The scaling dimension $\Delta_{12}$ is

\begin{equation}\label{scaling2}
 \Delta_{12}=\frac{1}{2}\left(F_1+{F_2}\right),
 \end{equation}
where, for repulsive interactions  
\begin{eqnarray}
 F_1&=&\sqrt{\frac{1+\lambda_4^1-\lambda_2^0-\lambda_2^1}{1+\lambda_4^1+\lambda_2^0+\lambda_2^1}},\quad F_1\in [0,1]\\ 
 F_2&=&\sqrt{\frac{1-\lambda_4^1+\lambda_2^0-\lambda_2^1}{1-\lambda_4^1-\lambda_2^0+\lambda_2^1}},\quad F_2\in [0,\infty]
\end{eqnarray}

\noindent with $\lambda_a^b=\frac{g_a^b}{\pi(v_F+g_4^0/\pi)}$. In the limit $g_4^0=g_2^0=g_4^1=0$ 
of no interactions within the helical modes the scaling dimension $\Delta_{12}=\sqrt{\frac{1-\lambda_2^1}{1+\lambda_2^1}}<1$   is the same as for  scattering by magnetic impurities \cite{Beri2012}.

\subsection{Random Disorder}

Now we switch to the case of multiple impurities on  the edge. This perturbation is described by 
\begin{equation}\label{dis_scat}
 \mathcal{O}^{\rm dis}_{ij}=\int dx \xi_{ij}(x)(\psi_{R,i}^\dagger\psi_{L,j}-\psi_{L,i}^\dagger\psi_{L,j}).
\end{equation}
Here $\xi_{ij}(x)$ is the (random) scattering amplitude, and $\xi_{ii}=0$ due to the TR symmetry.
%We now add these terms to the Hamiltonian and study whether they lead to Anderson localization.
We model the scattering to be local along the edge and uncorrelated for the different pairs of helical states   
\begin{equation}
\langle\xi_{ij} (x)\xi_{kl}(x')\rangle=W_{ij}\delta_{ik}\delta_{jl}\delta(x-x').
\end{equation}
We now  follow the steps of Giamarchi-Schultz renormalization group analysis\cite{Giamarchi1988}.
For the weak disorder  one finds\cite{Giamarchi1988}
\begin{equation}\label{RG_eq}
 \frac{dW_{ij}}{d\ell}=(3-2\Delta_{ij})W_{ij},
\end{equation}

\noindent where $\Delta_{ij}=\Delta_{|i-j|}$ is the scaling dimension of  scattering process (\ref{dis_scat}) 
between helical states $i$ and $j$ allowed by TR symmetry.
%The scaling dimension of $\Psi_{ij}$ controls the decay of the expectation value $\langle\Psi_{ij}(0)\Psi_{ij}(x)\rangle \sim |x|^{-2\Delta_{ij}}$.
%This operator (\ref{dis_scat}) after bosonization becomes $\Psi_{ij}=e^{i(\varphi_{\uparrow,i}+\theta_{\uparrow,i}-\varphi_{\downarrow,j}+\theta_{\downarrow,j})}$. 
In the conducting phase the disorder is an irrelevant perturbation, and  all $W_{ij}$  flow to zero. 
This requires  $\Delta_{|i-j|}>\frac{3}{2}$ for all pairs $i,j$.
Let us consider  two limiting cases: (i)  disorder that mixes only  the nearest  modes 
$W_{ij}\sim W\delta_{i,j+1}$;
(ii) the disorder that mixes the modes uniformly  
$W_{ij}\sim W$. All physical realizations lie in between these two limits.
 
The simplest situation is realized for  $\nu=2$ where  the limits (i) and (ii)  coincide.
In that case, the scaling of the disorder operator is given by Eq.(\ref{scaling2}). In the absence of inter-mode
interaction  the scaling dimension of a back scattering operator is $\Delta_{12}=(K+K^{-1})/2$. Therefore the system  
flows to a  conducting fixed point for  $K<(3-\sqrt{5})/2$.
\begin{figure}[ht!]
\includegraphics[width=1\columnwidth]{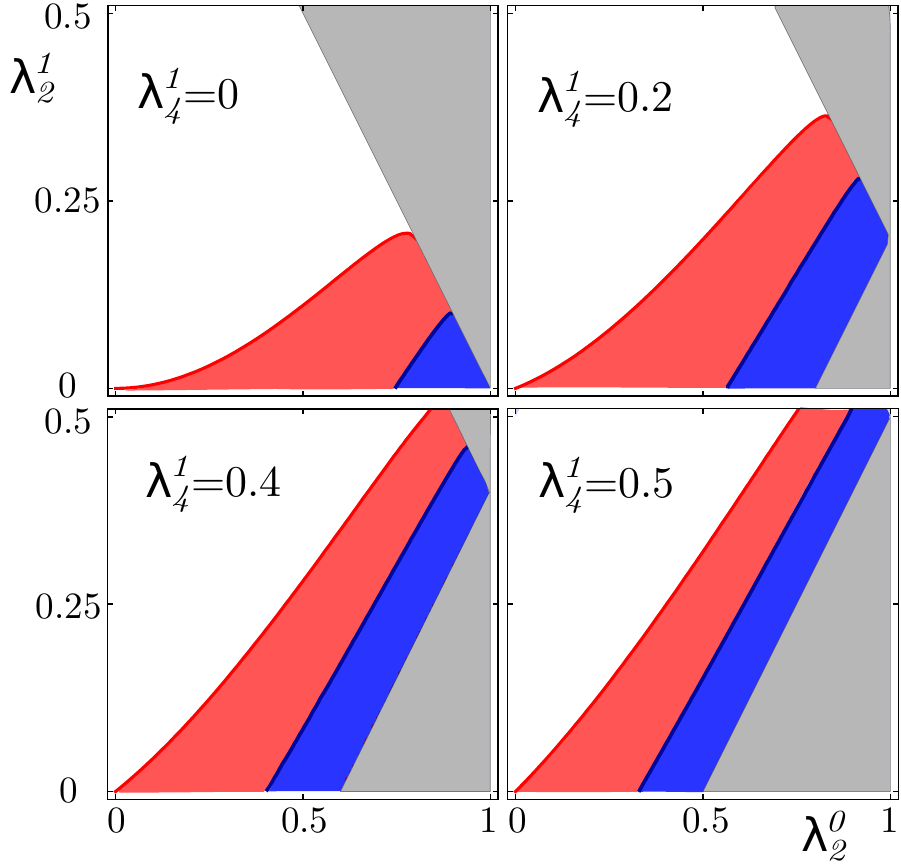}
\caption{\label{fig:Lambda} (color online) Phase diagram for  $(\lambda_2^0,\lambda_2^1)$.
Red region corresponds to the conducting phase for a single impurity.
Multiple impurities  are irrelevant in the blue region.
The gray region is forbidden by positivity of matrix $\mathcal{M}$.}
\end{figure}
In the presence of inter mode interaction the phase diagram  is show in Fig. (\ref{fig:Lambda})
as function of interaction parameters. 
The symmetry between intra and inter mode forward scattering ($\lambda_4^1\rightarrow 1$) 
enhances the conducting phase.

\subsection{Effect of two particle processes}

For weak interactions, two particle processes are less relevant than single particle events. For sufficiently strong
interactions, they start to compete. We analyze here the following processes involving two particle events
\begin{eqnarray}
 \mathcal{O}_{II,c}=t_c\int dx\delta(x)\psi_{R,1}^\dagger\psi_{L,1}^\dagger\psi_{R,2}\psi_{L,2},\\
 \mathcal{O}_{II,s}=t_s\int dx\delta(x)\psi_{R,1}^\dagger\psi_{R,2}^\dagger\psi_{L,1}\psi_{L,2},
\end{eqnarray}
which correspond to the transfer of $2e$ charge and two particle backscattering respectively. These processes renormalize 
according to 
\begin{equation}
 \frac{dt_a}{dl}=(1-\Delta_a)t_a,
\end{equation}
with $a=(s,c)$. Here $\Delta_a$ is the scaling dimension of the operator $\mathcal{O}_{II,a}$. They are
\begin{equation}\label{scaling_dim}
 \Delta_c=2F_2\quad\mbox{and}\quad\Delta_s=2F_1.
\end{equation}

\begin{figure}[ht!]
\includegraphics[width=1\columnwidth]{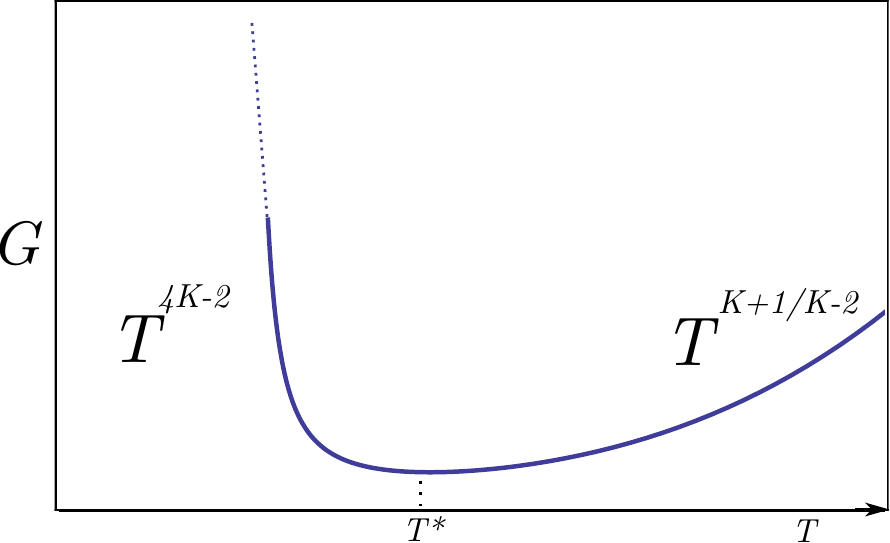}
\caption{\label{fig:non-mon} (color online).  Sketch of the conductance as function
of temperature. For $F_1<1/2$,  the two particle process $\mathcal{O}_{II,s}$ becomes relevant,
leading to a non monotonous behaviour of the conductance at high temperatures. In the figure we take
$F_1=K$ and $F_2=1/K$, which correspond to the simplest case of interactions just within each helical mode, 
discussed in Eq. (\ref{Ham_simple})}
\end{figure}

In the case of $1/2<F_1<1$, the system is in the conducting phase for $F_1+F_2>2$. The correction to the conductance $G=4e^2/h$ scales with 
the temperature as
\begin{equation}
 \delta G\sim  \begin{cases} -c_1\mu_{12}\left(\frac{2\pi a T}{u}\right)^{F_1+F_2-2}, & \quad \text{if } F_2 < 3F_1,\\ -c_2\mu_{12}^2v_F\left(\frac{2\pi a T}{u}\right)^{4F_1-2}, & \quad \text{if } F_2>3F_1.\\ \end{cases} 
\end{equation}
where $c_i$ are non-universal parameters. If $F_1<1/2$, the second order process $\mathcal{O}_{II,s}$ becomes relevant. The
conductance then becomes non monotonous at large temperatures (see  Fig.\ref{fig:non-mon})

For random disorder, the two particle operators are
\begin{eqnarray}
 \mathcal{D}_{II,c}=\int dx\xi_c(x)\psi_{R,1}^\dagger\psi_{L,1}^\dagger\psi_{R,2}\psi_{L,2},\\
 \mathcal{D}_{II,s}=\int dx\xi_s(x)\psi_{R,1}^\dagger\psi_{R,2}^\dagger\psi_{L,1}\psi_{L,2},
\end{eqnarray}
where the  $\xi(x)_a$ are uncorrelated random variables with $\langle\xi_a(x)\xi_b(x')\rangle=W_a\delta_{ab}\delta_(x-x')$.
These processes renormalize acording to the RG equations
\begin{equation}
 \frac{dW_a}{dl}=\left(\frac{3}{2}-\Delta_{a}\right)W_a,
\end{equation}
with $\Delta_{a}$ given by (\ref{scaling_dim}). For $3/4<F_1<1$ and $F_1+F_2>3$, the conducting phase remains.
If $F_1<3/4$ (still with $F_1+F_2>3$), the process $\mathcal{D}_{II,s}$ becomes relevant under RG and conductance
becomes non-monotonous at high temperatures (similar to the case of single impurity).

\subsection{$\nu\gg1$ Helical Edge Modes}

We now proceed with a more  general case of $\nu$ helical states.
To calculate  $\Delta$,  we consider the  operators of the form $\Psi_{\boldsymbol{m}}=e^{i\boldsymbol{m}\cdot\boldsymbol{\Phi}}$
where each vector $\boldsymbol{m}=(\boldsymbol{m}_R,\boldsymbol{m}_L)$ 
corresponds to a different  physical process. For example 

\begin{equation}\label{m_l_dist}
 (\boldsymbol{m}_R)_i=\delta_{ik},\quad  (\boldsymbol{m}_L)_i=-\delta_{i,k+l},
\end{equation}

\noindent describes an operator $\Psi_{\boldsymbol{m}}$ that backscatter a right mover in the mode $k$ to a left mover in the 
mode $k+l$. Using the (quadratic) action (\ref{action}),
one computes  the scaling dimension of $\Psi_{\boldsymbol{m}}$ 

\begin{equation}\label{Delta}
 \Delta[\Psi_{\boldsymbol{m}}]=\frac{1}{2}\boldsymbol{m}^T\varLambda\boldsymbol{m},
\end{equation}

\noindent where $\varLambda =\mathcal{M}^{-\frac{1}{2}}|\mathcal{M}^{\frac{1}{2}}\mathcal{K}\mathcal{M}^{\frac{1}{2}}|\mathcal{M}^{-\frac{1}{2}}.$
Here the absolute value of a matrix in the right hand side is defined as the absolute value of its eigenvalues. In other words, if $A$ is a diagonalizable matrix $A=UDU^{-1}$, then 

\begin{equation}
 |A|=U|D|U^{-1}=U
 \begin{pmatrix}
  |d_1| &0&\cdots\\
  0 &|d_2|& \cdots\\
  \vdots& & \ddots
 \end{pmatrix}U^{-1}.
\end{equation}

Note that, for translationally invariant interaction we consider,  
 the interaction matrices $M$ are of Toeplitz type, i.e.
$(M_{\rm fw/bw})_{ij}=(M_{\rm fw/bw})_{|i-j|}$.

Now on  we focus on  the limit where the  number of modes is large ($\nu \gg1 $).
Therefore,  one  can impose the periodic boundary condition in the mode space, 
without  changing  the results. 
In this case the interaction matrices are circulant,  $M_{|\nu-i-j|}=M_{|i-j|}$, and can be easily diagonalized
\cite{Circulant}.

 We adopt a  $g$-ology type notations  and model the interaction 
by $g_2$ and $g_4$ components
\begin{eqnarray}\label{int_same}
 (M_{\rm fw})_{|i-j|}&=&v_F\delta_{ij}+g_4(|i-j|),\\
 (M_{\rm bw})_{|i-j|}&=&-g_2(|i-j|).
\end{eqnarray}

Here  the distance dependent $g_4(i)$ accounts for the forward  interactions between electron densities
of the same chirality at distance $i$, while $g_2(i)$ parameterizes the backward interactions of densities of opposite chiralities.
The scaling dimension $\Delta_\ell$,  defined in  Eq.(\ref{Delta}) with 
$\boldsymbol{m}$ given by Eq.(\ref{m_l_dist}) is

\begin{equation}\label{Deltasame}
 \Delta_\ell=\frac{1}{\nu}\sum_{k=0}^{\nu-1}\frac{1-G(k)\cos(2\pi k \ell/\nu)}{\sqrt{1-G^2(k)}}.
\end{equation}
The  function $G(k)$,  is determined by the interaction parameters $g_{2,4}$ 
\begin{equation}
 G(k)=\frac{\tilde{g_2}(k)}{v_F+\tilde{g_4}(k)}.
\end{equation}

\noindent Here $\tilde{g}_{2,4}(k)=\sum_{j=1}^\nu \cos(2\pi jk/\nu)g_{2,4}(j)$ is the cosine transform of ${g}_{2,4}(r)$. The condition of $|G(k)|< 1$ follows from 
positivity of matrix $\mathcal{M}$. With the scaling of disorder operators at hand we can analyze their behavior under renormalization.

We now focus on   a finite range interaction.
One can easily show that the scattering processes between distant modes 
are less effective for the localization than backscattering between close  ones.
For the model of isotropic interaction 
$g_4(r)=g_2(r)=g\exp( - r^2/R^2)$ 
the scattering between the distant modes (\ref{dis_scat}) 
is  irrelevant for
\begin{equation}\label{condition}
 \sqrt{\frac{g}{2v_F}} > \frac{1}{\pi}\frac{R}{d}.
\end{equation}
%for $r<R$, and $g(r)=0$ for $r>R$, 
Here we  assumed  that  $|i-j|=\ell \gg 1$
and  $g/v_F \gg 1$.

\begin{figure}[ht!]
\includegraphics[width=1\columnwidth]{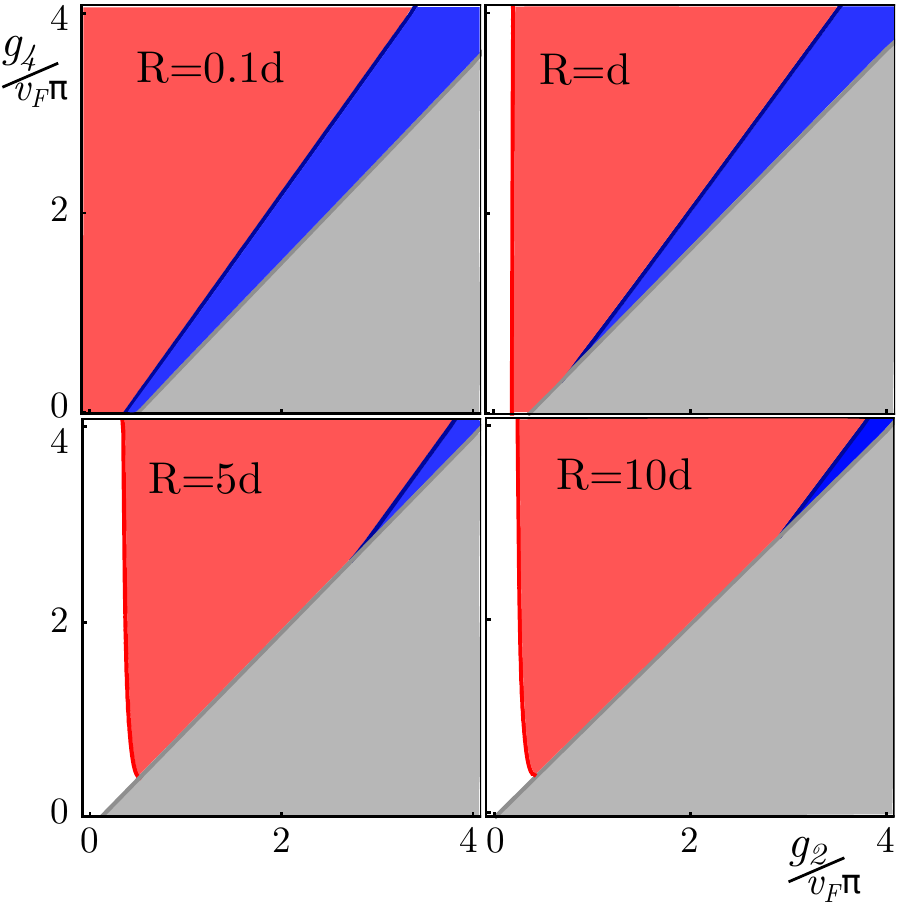}
\caption{\label{fig:l2} (color online).  $(g_2,g_4)$ phase diagram $\nu \gg 1$,  
finite ranges interaction $g_{2,4}(n)=g_{2,4}\exp\left(-(\frac{nd}{R})^2\right)$. 
Panels correspond to the  different values of the interaction radius $R$.
Color code is the same as in Fig.\ref{fig:Lambda}}
\end{figure}

The scattering  between the nearest modes ($\ell =1$)  imposes more stringent conditions
on the interaction constants $\lambda_a^b$, as shown in Fig.\ref{fig:l2}.
In particular, the conducting phase is stable only for the nearly symmetric interaction.  
For the fixed values of interaction strength the localization is enhanced 
by  increasing the interaction radius. 
In other words,  strong and short range interaction most efficiently 
drives the system towards the conducting phase.

\section{Summary}

To summarize, we have studied the localization of   the edge modes in  TIs with TR symmetry.
 We find that a combination of TR symmetry  and zero bias anomaly  changes the scaling dimensions of scattering operators.  This notably affects the phase diagram.  
For a sufficiently  strong  values of interaction the  zero temperature fixed point
is a conductor with  a number of  edge modes that are  stable against TR disorder. 
This  holds also  for the  even fillings, where the non-interacting system is equivalent to a trivial insulator.
We have analyzed the problem in several limiting cases, for the 
single impurity and  random disorder,
short and long range interaction, for  a variety of filling fractions $\nu$.
We have computed  the boundaries of the conducting phase in all these cases.
For intermediate values of interaction  electric conductivity
is a non-monotonous function of  temperature, due to interplay of single and two electron scattering processes.

\medskip

The authors acknowledge discussions with E. Berg,  Y. Gefen,  I. V.Gornyi, N. Kainaris,
A.D. Mirlin, I.V.  Protopopov, E. Sela.
This work was  supported by  GIF  and ISF.

\end{document}